\documentclass[11pt,dvips]{article}

\usepackage{epsfig}
\usepackage{graphicx,float}
\usepackage{array}
\usepackage{tabularx}
\usepackage{indentfirst}
\usepackage{amsmath,amssymb,bbold}
\usepackage{url}
\usepackage{rotating}
\usepackage{color}
\usepackage{hhline}
\usepackage{hyperref}
\usepackage{slashed}
\usepackage{tikz}
\usepackage{makecell}
\usepackage{pgfplots}
\usepackage{empheq}
\usepackage{physics}

\newcommand{\s}{\\ \vspace*{-3.5mm}}
 \newcommand{\rb}[2]{\raisebox{#1}[-#1]{#2}}

\renewcommand{\thefootnote}{\alph{footnote}}

\begin{document}


\begin{flushright}
\end{flushright}

\vskip 1.cm

\begin{center}
{\LARGE \bf Selection rules for the decay of a particle into \\[1mm]
            two identical massless particles of any spin}\\[1.0cm]
{Seong Youl Choi\footnote{sychoi@jbnu.ac.kr} and
 Jae Hoon Jeong\footnote{jaehoonjeong229@gmail.com}} \\[0.5cm]
{\it  Department of Physics and RIPC,
      Jeonbuk National University, Jeonju 54896, Korea}
\end{center}

\vskip 0.5cm

\begin{abstract}
\noindent
The well-known Landau-Yang (LY) theorem on the decay of a neutral particle
into two photons is generalized for analyzing the decay of a neutral or
charged particle into two identical massless particles of any spin. 
Selection rules categorized by discrete parity invariance and Bose/Fermi 
symmetry are worked out in the helicity formulation. The general form of 
the Lorentz-covariant triple vertices are derived and the corresponding 
decay helicity amplitudes are explicitly calculated in the Jacob-Wick 
convention. After checking the consistency of all the analytic results 
obtained by two complementary approaches, we extract out the key aspects 
of the generalized LY theorem.
\end{abstract}



\vskip 1.cm

\renewcommand{\thefootnote}{\fnsymbol{footnote}}

\section{Introduction}
\label{sec:introduction}

In the Standard Model
(SM)~\cite{Glashow:1961tr,Weinberg:1967tq,Salam:1968rm,Fritzsch:1973pi},
the photon and gluons are spin-1 massless particles.  
The direct observation of gravitational waves~\cite{Abbott:2016blz,
TheLIGOScientific:2017qsa,LIGOScientific:2018mvr} strongly indicates
the presence of spin-2 massless bosons called gravitons
at the quantum level. On the other hand, it is not yet clear whether 
the lightest neutrino is massless or not~\cite{Zyla:2020zbs} and 
it is a hotly-debated issue whether there can exist any massless 
particle with its spin larger
than two~\cite{Weinberg:1980kq,Loebbert:2008zz,Bekaert:2010hw}.
In this situation, any conceptual and/or practical studies of massless 
particles with arbitrary spin are still of great interest. \s

More than seventy years ago, Landau~\cite{Landau:1948kw} and Yang~\cite{Yang:1950rg} have shown that selection rules governing the
decay of a particle into two photons, which are the prototype of
spin-1 massless particles, can be derived from the general
principle of invariance under rotation and inversion.
Concisely speaking, in terms of the number $n$ of possible two-photon states
for the $J$ and intrinsic parity $\eta$ of the decaying particle
can the selection
rules be summarized collectively with the compact notation $n\, [J]^{\eta}$ as
\begin{eqnarray}
{\large n\, [J]^{\eta}} \ \ = \ \ 1\,[0]^+\,, \ \ 2\,[2k]^+\,, \ \
             1\,[2k+1]^+ \quad  \mbox{or}\quad   1\,[0]^-\,,\ \ 1\, [2k]^-\,,
\label{eq:landau_yang_theorem}
\end{eqnarray}
with a positive integer $k$. One consequence of the
so-called Landau-Yang (LY) theorem is that {\it no on-shell spin-1
particle can decay into two on-shell massless photons.} [Therefore, both the
gluon-fusion production and the two-photon decay of the resonance with
mass of 125 GeV discovered and confirmed at the
LHC ~\cite{Aad:2012tfa,Chatrchyan:2012ufa}
exclude the possibility of the spin being unity. i.e. 
$J \neq 1$~\cite{Choi:2012yg}.]\s

In this letter, we generalize the LY theorem for investigating the decay
$X\to MM$ of a neutral or charged particle into two identical massless
particles of {\it any spin}. We note in passing that, based on the approaches
by Landau~\cite{Landau:1948kw}, the generalization of the LY theorem to
the case with spin-0 and spin-1/2 massless particles has been investigated
in Ref.$\,$\cite{Pleitez:2015cpa}. Besides, although not studied here,
we note that the LY theorem may be avoided if some of the basic assumptions
for the theorem are not imposed~\cite{Beenakker:2015mra,Cacciari:2015ela,
Pleitez:2018lct,Behr:2002wx,Weinberg:1984vb,Ivanov:2019lgh}.\s

Firstly, we work out the selection rules categorized by discrete parity
invariance and Bose/Fermi symmetry due to two identical bosons/fermions
on the decay in the helicity formulation based on the Jacob-Wick
convention~\cite{Jacob:1959at} in
Section~\ref{sec:selection_rules_helicity_formalism}. Secondly, we derive
the general form of the Lorentz covariant triple vertices explicitly in
Section~\ref{sec:general_form_lorentz_covariant_vertex} and then
calculate the corresponding decay helicity amplitudes in detail in
Section~\ref{sec:explicit_form_reduced_helicity_amplitudes}.\footnote{Another
convenient procedure for describing the $XMM$ triple vertex is
to use a spinor formalism developed for handling massive as well as
massless particles in Ref.~\cite{Arkani-Hamed:2017jhn}.}
After checking the consistency of all the analytic results obtained
by two complementary approaches, we summarize the key aspects of the
generalized LY theorem and conclude in Section~\ref{sec:conclusions}.
All the formulae useful for explicitly deriving the decay helicity
amplitudes in the main text are listed in
Appendices~\ref{appendix:wave_vectors_spinors_jacob_wick} and
\ref{appendix:scalar_vector_tensor_currents_jacob_wick}.\s

\section{Selection rules in the helicity formalism}
\label{sec:selection_rules_helicity_formalism}

\begin{figure}[H]
\begin{center}
\includegraphics[width=8.5cm, height=5.5cm]{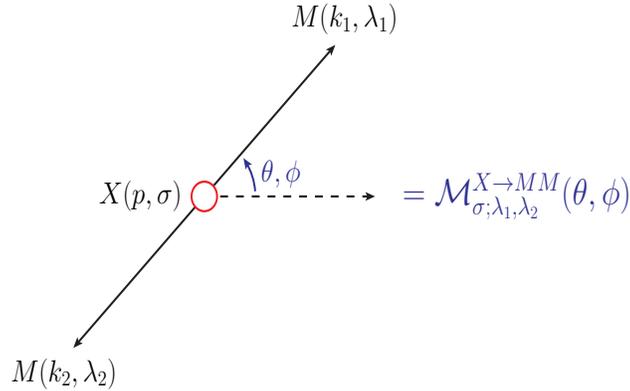}
\caption{\it Kinematic configuration for the helicity amplitude of the two-body
             decay $X\to MM$ of $X$ into two identical massless
             particles $MM$ in the $X$ rest frame ($X$RF). The notations,
             $\{p, k_{1,2}\}$ and $\{\sigma, \lambda_{1,2}\}$, are the momenta
             and helicities of the decaying particle $X$ and two massless
             particles $MM$, respectively. The polar and azimuthal
             angles, $\theta$ and $\phi$, are defined with respect to an
             appropriately chosen coordinate system.
}
\label{fig:kinematic_configuration_xmm_xrf}
\end{center}
\end{figure}

The helicity formalism~\cite{Jacob:1959at,Haber:1994pe} is one of the most
efficient tools for discussing the two-body decay of a spin-$J$ particle $X$
into two of a massless particle $M$ of spin $s$,
treating any massive and massless particles on an equal footing.  
For the sake of a transparent and straightforward analytic analysis, 
we describe the two-body decay $X\to MM$ in the $X$ rest frame ($X$RF)
\begin{eqnarray}
X(p,\sigma)
   \ \ \rightarrow\ \
M(k_1,\lambda_1)\, +\, M(k_2,\lambda_2)\,,
\label{eq:x_mm_decay}
\end{eqnarray}
in terms of the momenta, $\{p, k_1, k_2\}$, and 
helicities, $\{\sigma, \lambda_1, \lambda_2\}$, of the particles, 
as depicted in Figure~\ref{fig:kinematic_configuration_xmm_xrf}.\s

\vskip 1.cm

Before going into a detailed description of the general Lorentz-covariant form
$XMM$ vertex in Section~\ref{sec:general_form_lorentz_covariant_vertex},
we study general restrictions on the decay helicity amplitude due to space
inversion and Bose/Fermi symmetry for two identical massless
bosons/fermions in the final state.\s

The helicity amplitude of the decay $X\to MM$ is decomposed in terms
of the polar and azimuthal angles, $\theta$ and $\phi$, defining the
direction of one massless particle in a fixed coordinate system
as
\begin{eqnarray}
  {\cal M}^{X\to MM}_{\sigma;\lambda_1,\lambda_2}(\theta,\phi)
\,\, =\,\, {\cal C}_{\lambda_1,\lambda_2} \,\,
  d^{J}_{\sigma,\,\lambda_1-\lambda_2}(\theta)\,
  e^{i(\sigma-\lambda_1+\lambda_2)\phi}
  \quad\ \ \mbox{with}\quad\ \
  |\lambda_1-\lambda_2|\leq J\,.
\label{eq:x_mm_helicity_amplitude}
\end{eqnarray}
with the constraint $|\lambda_1-\lambda_2|\leq J$ in the Jacob-Wick
convention~\cite{Jacob:1959at,Haber:1994pe} (see
Figure~\ref{fig:kinematic_configuration_xmm_xrf} for the kinematic configuration).
Here, the helicity $\sigma$ of the spin-$J$ massive particle $X$ takes
one of $2J+1$ values between $-J$ and $J$. In contrast, the helicities,
$\lambda_{1,2}$, can take only two values $\pm s$ because only the
maximal-magnitude helicity values are allowed for any massless particle.
Therefore, there exist {\it at most
four independent $(\lambda_1,\lambda_2)$ combinations $(\pm s, \pm s)$ and
$(\pm s, \mp s)$}, to be denoted by the shortened notations $(\pm, \pm)$
and $(\pm, \mp)$ in the following. We note that the reduced helicity
amplitudes ${\cal C}_{\lambda_1,\lambda_2}$ do not depend on the $X$ helicity
$\sigma$ due to rotational invariance and the polar-angle dependence is 
fully encoded
in the Wigner $d$ function $d^{J}_{\sigma,\lambda_1-\lambda_1}(\theta)$ given
in the convention of Rose~\cite{merose2011}.\s

\vskip 0.5cm

\begin{figure}[H]
\begin{center}
\includegraphics[width=8.5cm, height=5.5cm]{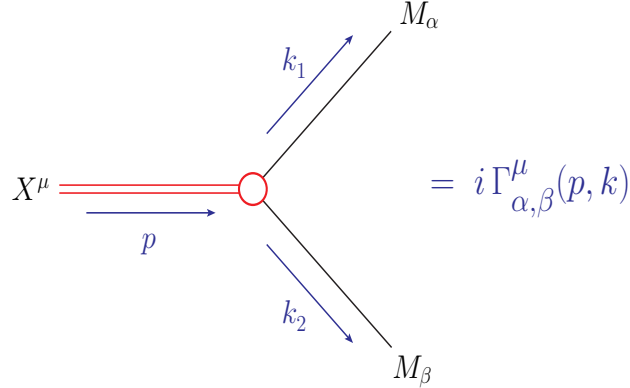}
\caption{\it Feynman rules for the general $XMM$ triple vertex of a spin-$J$
             particle $X$ and two identical spin-$s$ massless
             particles $MM$. The indices, $\mu$, $\alpha$ and $\beta$, stand
             for the sequences of $\mu=\mu_1\cdots \mu_{_J}$,\,
             $\alpha_1\cdots \alpha_{s}$ and $\beta_1\cdots\beta_{s}$,
             collectively. The symmetric and anti-symmetric momentum
             combinations, $p=k_1+k_2$ and $k=k_1-k_2$, are introduced
             for systematic classifications of the triple vertex tensor.
}
\label{fig:xmm_vertex}
\end{center}
\end{figure}

Bose or Fermi symmetry for the two identical integer or half-integer spin
particles in the final state leads to the relation for the reduced decay
helicity amplitudes:
\begin{eqnarray}
  {\cal C}_{\lambda_1,\lambda_2}
\, \, =\, \, (-1)^{J}\,\, {\cal C}_{\lambda_2,\lambda_1}
\ \ \ \ \mbox{with}\ \ \ \
  |\lambda_1-\lambda_2|\leq J\,,
\label{eq:x_mm_identical_condition}
\end{eqnarray}
derived by the (anti)-symmetrization of the final state of two identical
particles. If the decay process conserves parity, the reduced helicity
amplitudes satisfy the space-inversion or parity relation
\begin{eqnarray}
    {\cal C}_{\lambda_1,\lambda_2}
= \eta (-1)^{J-2s} \,\, {\cal C}_{-\lambda_1,-\lambda_2}
\ \ \ \ \mbox{with}\ \ \ \
  |\lambda_1-\lambda_2|\leq J\,,
\label{eq:x_mm_parity_relation}
\end{eqnarray}
with $\eta$ being the intrinsic $X$ parity. Once more, we emphasize that
there are only four helicity combinations $(\pm, \pm)$ and $(\pm, \mp)$ for
the final-state system of two massless particles. \s

Remarkably, the identical-particle (ID)
condition (\ref{eq:x_mm_identical_condition})
and parity (PA) relation (\ref{eq:x_mm_parity_relation}) enable us to
straightforwardly derive the selection/exclusion rules for the two-body
decay $X\to MM$ classified according to the spin $J$, the spin $s$,
the $X$ intrinsic parity $\eta$ and also whether $J < 2s$ or not.
The opposite-helicity amplitudes,
${\cal C}_{\pm,\mp}$, vanish for $J<2s$ and the same-helicity amplitudes,
${\cal C}_{\pm,\pm}$, vanish for any odd integer $J$. Consequently,
{\it one important exclusion rule is that any decay with odd $J$ less than
$2s$ is forbidden irrespective of the $X$ parity.} Specifically, the
well-known rule that any spin-1 particle cannot decay into two identical
massless spin-1 particles such as photons and color-neutral gluons can be
guaranteed because $J=1$ is odd and $J=1 < 2 s = 2$ for
$s=1$. Moreover, for $J\geq 2s$, the same-helicity amplitudes
still vanish for any odd $J$. According to the ID condition (\ref{eq:x_mm_identical_condition}) and the PA
relation (\ref{eq:x_mm_parity_relation}), the opposite-helicity
amplitudes survive only when $\eta\,(-1)^{2s}=+1$ with the relation
${\cal C}_{+,-}=(-1)^J\, {\cal C}_{-,+}$.\s

Based on the above observations, a few interesting selection rules can be
extracted out as follows:
\begin{itemize}
\item No odd-$J$ particle can decay into two identical spin-0 massless
      scalar bosons ($s=0$).
\item The allowed decay of a spin-1 massive particle into two identical
      massless particles is only into two identical spin-1/2 massless 
      fermions, when the intrinsic parity $\eta$ of the decaying particle
      is odd.
\item Any even-parity and odd-spin particle cannot decay into two
      identical massless fermions.
\item Any odd-parity and odd-spin particle cannot decay into two
      identical massless bosons.
\end{itemize}
Table~\ref{tab:selection_rules} summarizes all the selection/exclusion rules
on the special decay $X\to MM$. We note that the case with $s=0$ is more
restricted than the case with non-zero spin $s$, as marked with the
comment [forbidden].\s

Interestingly, as pointed out in Ref.$\,$\cite{Arkani-Hamed:2017jhn}, 
the general selection rules enable us to re-interpret the Weinberg-Witten (WW)
theorem~\cite{Weinberg:1980kq}.
Since conserved currents and stress tensors measure the charge and momentum
on single particle states respectively, let us consider the decay of
a massive neutral particle into two {\it opposite-helicity} massless particles,
i.e. $\lambda_1=-\lambda_2=\pm$, because two momenta $k_1$ and $k_2$ are
out going and so $\lambda_1=-\lambda_2$ in order for $k_1$ and $k_2$
to represent the same particle. In this case, as can be checked with Table~\ref{tab:selection_rules}, the inequality $s\leq J/2$ must be
satisfied, i.e. $s\leq 1/2$ for $J=1$ and $s\leq 1$
for $J=2$. That is to say, massless particles
with spin $s> 1/2$ cannot couple to a Lorentz covariant conserved
current corresponding to $J=1$ and those with spin $s>1$ cannot couple
to a conserved energy-momentum stress tensor corresponding to $J=2$. \s

\vskip 0.5cm

\begin{table}[H]
\centering
\begin{tabular}{||c||c|c|c|c||c|c|c|c||} \hline\hline
& \multicolumn{4}{c||}{ }
& \multicolumn{4}{c||}{ } \\[-2mm]
{\color{black} \boldmath{$J$}}
& \multicolumn{4}{c||}{\color{blue} $J < 2s$}
& \multicolumn{4}{c||}{\color{blue} $J\geq 2s\,\, [s=0]$}
  \\[2mm] \hline\hline
& \multicolumn{4}{c||}{ }
& \multicolumn{2}{c|}{ }
& \multicolumn{2}{c||}{ }
  \\[-3mm]
& \multicolumn{4}{c||}{\small $\zeta=\pm $}
& \multicolumn{2}{c|}{\small $\zeta=+ $}
& \multicolumn{2}{c||}{\small $\zeta=- $}
  \\[2mm] \cline{2-9}
& \multicolumn{4}{c||}{ }
& \multicolumn{2}{c|}{ }
& \multicolumn{2}{c||}{ }
  \\[-2mm]
 \rb{2.5ex}{\color{blue} odd}
& \multicolumn{4}{c||}{\color{black}\small forbidden}
& \multicolumn{2}{c|}{\color{black}\small forbidden}
& \multicolumn{2}{c||}{\color{black} ${\cal C}_{+,-}\, =\, -{\cal C}_{-,+}$\,
                                      {\small [forbidden]}}
  \\[2mm] \hline\hline
& \multicolumn{4}{c||}{ }
& \multicolumn{2}{c|}{ }
& \multicolumn{2}{c||}{ }
  \\[-2mm]
& \multicolumn{4}{c||}{\small $\zeta=\pm$}
& \multicolumn{2}{c|}{\small $\zeta=+ $}
& \multicolumn{2}{c||}{\small $\zeta=- $}
  \\[2mm] \cline{2-9}
& \multicolumn{4}{c||}{ }
& \multicolumn{2}{c|}{ }
& \multicolumn{2}{c||}{ }
  \\[-2mm]
 \rb{2.5ex}{\color{blue} even}
& \multicolumn{4}{c||}{\color{black} ${\cal C}_{+,+}\, =\, \pm {\cal C}_{-,-}$}
& \multicolumn{2}{c|}{{\color{black} ${\cal C}_{+,-}= {\cal C}_{-,+}$\,
                                      \small [forbidden]\,,\,\,
                                      ${\cal C}_{+,+}\, =\, {\cal C}_{-,-}$
                      }}
& \multicolumn{2}{c||}{\color{black} ${\cal C}_{+,+}\, =\, -{\cal C}_{-,-}$\,
                                      \small [forbidden]}
  \\[2mm] \hline\hline
\end{tabular}
\vskip 0.3cm
\caption{\it Selection rules for the decay $X\to MM$. The values, $J$ and $s$,
are the spins of the massive $X$ and massless $M$ particles, respectively,
and the signature $\pm$ denotes the sign of the product 
$\zeta\equiv \eta (-1)^{J-2s}$ of the
$X$ intrinsic parity $\eta$ and the factor $(-1)^{J-2s}$. The mark $[fobidden]$
means that the corresponding term is absent when $s=0$.}
\label{tab:selection_rules}
\end{table}

In the next section, we show how the selection rules derived in the
helicity formalism are reflected and encoded in the Lorentz-covariant
form of the triple $XMM$ vertex in the integer and half-integer $s$
cases separately. \s

\section{General Lorentz-covariant triple vertex tensors}
\label{sec:general_form_lorentz_covariant_vertex}

Generically, the decay amplitude of one on-shell particle $X$ of mass
$m$ and integer spin $J$ into two of a massless particle $M$ of spin
$s$ can be written in terms of the triple vertex tensor $\Gamma$ as
\begin{eqnarray}
    {\cal M}^{X\to MM}_{\sigma; \lambda_1,\lambda_2}
&=& \bar{u}_1^{\alpha_1\cdots\alpha_{s}}(k_1,\lambda_1)\,\,
    \Gamma^{\mu_1\cdots\mu_J}_{\alpha_1\cdots\alpha_s,\beta_1\cdots\beta_s}(p,k)
     \,\, v_2^{\beta_1\cdots\beta_{s}}(k_2,\lambda_2)\,\,
    \epsilon_{\mu_1\cdots\mu_J}(p,\sigma)\,,
\label{eq:xmm_vertex}
\end{eqnarray}
with $\lambda_{1,2}=\pm s=\pm $ for the two-body decay $X\to MM$, where $p$ and
$\sigma$ are the momentum and helicity of the particle $X$, and $k_{1,2}$
and  $\lambda_{1,2}$ are the momenta and helicities of two massless particles,
respectively. Here, $p=k_1+k_2$ and $k=k_1-k_2$, are symmetric and
anti-symmetric under the interchange of two momenta, $k_1$ and $k_2$.
If $s$ is an integer, then the wave tensors
$\bar{u}_1^{\alpha_1\cdots\alpha_2}(k_1,\lambda_1)$
and $v_2^{\beta_1\cdots\beta_s}(k_2,\lambda_2)$ are given by
\begin{eqnarray}
       \bar{u}_1^{\alpha_1\cdots\alpha_s}(k_1,\pm s)
\, &=& \, \epsilon_1^{*\alpha_1\cdots\alpha_s}(k_1,\pm s)\,, \\
       v_2^{\beta_1\cdots\beta_s}(k_2,\pm s)
\, &=& \, \epsilon_2^{*\beta_1\cdots\beta_s}(k_2,\pm s)\,,
\end{eqnarray}
and if $s=n+1/2$ is a half-integer, the wave tensors are given by
\begin{eqnarray}
       \bar{u}_1^{\alpha_1\cdots\alpha_s}(k_1,\pm s)
\, &=& \, \epsilon_1^{*\alpha_1\cdots\alpha_n}(k_1,\pm n)\, \bar{u}_1(k_1,\pm)\,,
       \\
       v_2^{\beta_1\cdots\beta_s}(k_2,\pm s)
\, &=& \, \epsilon_2^{*\beta_1\cdots\beta_n}(k_2,\pm n)\, v_2(k_2, \pm)\,,
\end{eqnarray}
where $\bar{u}_1(k_1,\pm)=u^\dagger_1(k_1,\pm)\gamma^0$ and the $u_1(k_1,\pm)$
and $v_2(k_2,\pm)$ are the spin-1/2 particle $u_1$ and anti-particle $v_2$ spinors. (See Appendix~\ref{appendix:wave_vectors_spinors_jacob_wick}
for their detailed expressions.)\s

An on-shell boson of integer spin $J$, non-zero mass $m$, momentum $p$
and helicity $\sigma$ is defined by a rank-$s$ tensor $\epsilon_{\mu_1\cdots\mu_{_J}}(p,\sigma)$~\cite{Behrends:1957rup,
Auvil:1966eao,Weinberg:1995mt} that is completely
symmetric, traceless and divergence-free
 \begin{eqnarray}
      \varepsilon^{\alpha\beta\mu_i\mu_j}\,
      \epsilon_{\mu_1\cdots\mu_i\cdots\mu_j\cdots\mu_{_J}}(p,\sigma)
&=& 0\,,
   \label{eq:totally_symmetric}\\[1mm]
      g^{\mu_i\mu_j}\,
      \epsilon_{\mu_1\cdots\mu_i\cdots\mu_j\cdots\mu_{_J}}(p,\sigma)
&=& 0\,,
   \label{eq:traceless}\\[1mm]
      p^{\mu_i}\,
      \epsilon_{\mu_1\cdots\mu_i\cdots\mu_{_J}}(p,\sigma)
&=& 0\,,
   \label{eq:divergence_free}
 \end{eqnarray}
and it satisfies the on-shell wave equation $(p^2-m^2)\,
\epsilon_{\mu_1\cdots\mu_{_J}}(p,\sigma)=0$ for any helicity value of
$\sigma$ taking an integer between $-J$ and $J$.
The wave tensor can be expressed explicitly by a linear combination of
$s$ products of spin-1 wave vectors with appropriate Clebsch-Gordon
coefficients. \s

\subsection{Massless particles of integer spin}
\label{subsec:integer_spin_case}

The wave tensor of a  massless particle of integer spin $s$ is given
simply by a product of $s$ spin-1 wave vectors, each of which carries
the same helicity of $\pm 1$, as
\begin{eqnarray}
      \epsilon^{\alpha_1\cdots\alpha_s}(k,\pm s)
\ \ = \ \
      \epsilon^{\alpha_1}(k,\pm)\cdots
      \epsilon^{\alpha_s}(k,\pm)\,,
\label{eq:massless_wave_tensor}
\end{eqnarray}
where, for notational convenience, the notation $\pm$ is used for $\pm 1$.
The wave tensor (\ref{eq:massless_wave_tensor}) also is completely
symmetric, traceless and divergence-free and it satisfies the wave
equation $k^2 \epsilon^{\alpha_1\cdots\alpha_s}(k, \lambda)=0$
with $k^2=0$ automatically.\s

To begin with, let us consider as a special case the decay $X\to \gamma\gamma$
of a massive integer-spin boson $X$ into two massless spin-1 photons, originally
investigated by Landau~\cite{Landau:1948kw}
and Yang~\cite{Yang:1950rg}.
Imposing the on-shell conditions valid for two spin-1 massless
photons~\footnote{While the on-shell conditions (\ref{eq:two_photons}) are
maintained, the wave vector $\epsilon^\mu_i$ of a spin-1 massless particle
can be adjusted by adding any term proportional to $k^\mu_i$, guaranteeing
that its spatial part orthogonal to $\vec{k}_i$ remains as two physical degrees
of freedom.}
\begin{eqnarray}
&& k_i\cdot \epsilon_i(k_i,\lambda_i)=0\quad \mbox{and} \quad
   k^2_i=0\quad [i=1,2]\,,
\label{eq:two_photons}
\end{eqnarray}
and performing the Bose symmetrization of two identical $\gamma$ states
in the final state, we can write the general $X\gamma\gamma$ vertex in
a greatly-simplified form~\cite{Choi:2021zvz} as
\begin{eqnarray}
   \Gamma^{\tiny\mbox{$X\to \gamma\gamma$}}_{\mu;\, \alpha,\beta}(p,k)
&=& \eta_+\,
    x^+_b\,\, g_{{_\bot} \alpha\beta}\,\, k_{\mu_1}\cdots k_{\mu_{_J}}/m^J
    \nonumber\\
&+& \eta_+\,
    x^-_b\,\, \imath\,\langle \alpha\beta p k\rangle\,\,
               k_{\mu_1}\cdots k_{\mu_{_J}}/m^{J+2}
    \nonumber\\
&+& \eta_+\,
    y^+_b\,\, \left(\, g_{{_\bot}\alpha\mu_1} g_{{_\bot}\beta\mu_2}
                           + g_{{_\bot}\beta\mu_1} g_{{_\bot}\alpha\mu_2}
                           - g_{_{\bot}\alpha\beta}
                            \, k_{\mu_1} k_{\mu_2}/m^2
                           \right)\,\,
                    k_{\mu_3}\cdots k_{\mu_{_J}}/m^{J-2}
    \nonumber\\
&+& \eta_-\,
    y^-_b\,\, \imath\, \left(\, g_{_{\bot}\alpha\mu_1}\langle \beta\mu_2 p k\rangle
                     + g_{_{\bot}\beta\mu_1}\langle \alpha\mu_2 p k\rangle
                          \, \right)\,\,
                    k_{\mu_3}\cdots k_{\mu_{_J}}/m^J\,,
\label{eq:general_x_gammagamma_coupling}
\end{eqnarray}
satisfying the orthogonality conditions, $k^\alpha_1\, \Gamma^{\tiny\mbox{$X\to \gamma\gamma$}}_{\mu;\, \alpha,\beta}(p,k)\,=\,
k^\beta_2\, \Gamma^{\tiny\mbox{$X\to
\gamma\gamma$}}_{\mu;\, \alpha,\beta}(p,k)\,=\,0$,
with $\mu$ denoting collectively $\mu_1\cdots\mu_{_J}$,
the projection factors, $\eta_\pm =[1\pm(-1)^{J}]/2$, and two momentum
combinations, $p=k_1+k_2$ and $k=k_1-k_2$, which are symmetric
and antisymmetric under the interchange of two massless particles,
i.e. $k_1\leftrightarrow k_2$ and $\alpha\leftrightarrow \beta$,
respectively. The antisymmetric tensor $\langle \alpha\beta pk\rangle
=\varepsilon_{\alpha\beta\rho\sigma} p^\rho k^\sigma $
in terms of the totally antisymmetric Levi-Civita tensor
with the sign convention $\varepsilon_{0123}=+1$.
For the sake of notation, the following orthogonal tensors are introduced,
\begin{eqnarray}
    g_{_{\bot}\alpha\beta}\,
&=& g_{\alpha\beta} - 2\, k_{2\alpha}k_{1\beta}/m^2\,,
    \nonumber\\
    g_{_{\bot}\alpha\mu_i}
&=& g_{\alpha\mu_i} - 2\, p_{\alpha}\, k_{1\mu_i}/m^2\,,
    \nonumber\\
    g_{_{\bot}\beta\mu_i}
&=& g_{\beta\mu_i} - 2\, p_{\beta}\, k_{2\mu_i}/m^2\,,
\label{eq:orthgonal_tensors}
\end{eqnarray}
with $i=1,\cdots, J$.
The totally symmetric wave tensor
$\epsilon^{\mu_1\cdots\mu_{_J}}(p,\sigma)$ to be coupled to the
triple vertex (\ref{eq:general_x_gammagamma_coupling})
guarantees the automatic symmetrization of all the $x^\pm_b$
and $y^\pm_b$ terms under any $\mu$-index permutations.
It is straightforward to derive the following
selection rules from the expression of the $X\gamma\gamma$ vertex in Eq.$\,$(\ref{eq:general_x_gammagamma_coupling}),
\begin{itemize}
\item The $Y^+_{1,2}$ terms survive for non-negative
      even integers, $J=0, 2, 4$, and so on.
\item The $Y^+_{3}$ term survives for positive
      even integers, $J=2,4$, and so on, satisfying $J\geq 2$.
\item The $Y^-_1$ term survives for positive odd
      integers, $J=3,5$, and so on, satisfying $J>2$.
\end{itemize}
One immediate consequence is that {\it any massive on-shell
spin-1 particle with $J=1$ cannot decay into two on-shell
identical massless spin-1 particles such as photons, because the
three $x^\pm_b$ and $y^+_b$ terms vanish with $\eta_+=0$ for $J=1$
and the $y^-_b$ term surviving only for odd $J$ contribute to
the vertex only for $J\geq 3$.}\s

For a massless boson $M=b$ of an arbitrary integer spin $s$,
the general Lorentz-covariant form of the $Xbb$ vertex can be written
compactly by introducing two scalar-type operators and two tensor-type
operators as
\begin{eqnarray}
    S^+_{\alpha_i\beta_i}
&=& g_{_{\bot}\alpha_i\beta_i}\,,
\label{eq:scalar_tensor_operators_1} \\
    S^-_{\alpha_i\beta_i}
&=& \imath\, \langle \alpha_i\beta_i p k\rangle/m^2\,,
\label{eq:scalar_tensor_operators_2} \\
    T^+_{\alpha_i\beta_i,\mu_{2i-1}\mu_{2j}}
&=& g_{_{\bot}\alpha_i\mu_{2i-1}}\, g_{_{\bot}\beta_i\,\mu_{2i}}
  + g_{_{\bot}\beta_i\mu_{2i-1}}\, g_{_{\bot}\alpha_i\,\mu_{2i}}
  - g_{_{\bot}\alpha_i\beta_i}\,
    k_{\mu_{2i-1}} k_{\mu_{2j}}/m^2\,,
\label{eq:scalar_tensor_operators_3}\\
    T^-_{\alpha_i\beta_i,\mu_{2i-1}\mu_{2i}}
&=& \imath\,(g_{_{\bot}\alpha_i\mu_{2i-1}}\,
    \langle \beta_i\, \mu_{2i} p k\rangle
  + g_{_{\bot}\beta_i\,\mu_{2i-1}}\,
    \langle \alpha_i \mu_{2i} p k\rangle)/m^2\,,
\label{eq:scalar_tensor_operators_4}
\end{eqnarray}
with $i=1,\cdots, s$. The general form of the vertex tensor
is then cast into a compact form as
\begin{eqnarray}
&&   \Gamma^{\tiny \mbox{$X\to bb$}}_{\mu;\, \alpha,\beta}(p,k)
 = \eta_+\, \left(x^+_b \,\, S^+_{\alpha_1\beta_1}
                 +x^-_b\,\, S^-_{\alpha_1 \beta_1} \right)\,
              S^+_{\alpha_2\beta_2}\, \cdots
              S^+_{\alpha_s\beta_s}\,\,
              k_{\mu_1}\,\cdots\, k_{\mu_{_J}}/m^J
    \nonumber\\
&& \mbox{ }\hskip 0.5cm  + \eta_+\,\theta(J-2s)\,\, y^+_b\,\,
     T^+_{\alpha_1\beta_1,\mu_1\mu_2}\, T^+_{\alpha_2\beta_2,\mu_3\mu_4}\,
      \cdots\, T^+_{\alpha_s\beta_s,\mu_{2s-1}\mu_{2s}}\,\,
             k_{\mu_{2s+1}}\,\cdots\, k_{\mu_{_J}}/m^{J-2s}
    \nonumber\\
&& \mbox{ }\hskip 0.5cm + \eta_-\,\theta(J-2s)\,\, y^-_b\,\,
     T^-_{\alpha_1\beta_1,\mu_1\mu_2}\, T^+_{\alpha_2\beta_2,\mu_3\mu_4}\, \cdots\, T^+_{\alpha_s\beta_s,\mu_{2s-1}\mu_{2s}}\,\, k_{\mu_{2s+1}}\cdots\, k_{\mu_{_J}}/m^{J-2s} \,,
\label{eq:general_x_bb_vertex_tensor}
\end{eqnarray}
with $\eta_\pm = [1\pm (-1)^J]/2$ and the step function $\theta(J-2s)=1$ and
$0$ for $J\geq 2s$ and $J< 2s$. Here, the indices, $\mu$, $\alpha$ and
$\beta$, stand collectively for $\mu=\mu_1\cdots \mu_{_J}$,
$\alpha=\alpha_1\cdots\alpha_s$ and $\beta=\beta_1\cdots\beta_s$.
One immediate consequence is that {\it the case with odd $J$ less than
$2s$ is forbidden irrespective of the $X$ intrinsic parity $\eta$}, which
is consistent with the corresponding selection rule listed in
Table~\ref{tab:selection_rules}. Specifically, any spin-1 particle cannot
decay into two identical spin-1 massless particles with $s=1$.\s

\subsection{Massless particles of half-integer spin}
\label{subsec:half-integer_spin_case}

An on-shell massless particle or antiparticle of a half-integer spin
$s=n+1/2$ $(n=0,1,2,\cdots)$ and momentum $k$ may be described by a product of a
$u$ or $v$ spinor and $n$ spin-1 wave vectors as
\begin{eqnarray}
   u^\alpha(k,\pm s)
\, &=&\, \epsilon^{\alpha_1}(k,\pm)\, \cdots\, \epsilon^{\alpha_n}(k,\pm)\,
   \, u(k,\pm)\,, \\
   v^\alpha(k,\pm s)
\, &=&\, \epsilon^{*\alpha_1}(k,\pm)\cdots\epsilon^{*\alpha_n}(k,\pm)\,
   v(k,\pm)\,,
\label{eq:half-integer_spin_wave_tensor}
\end{eqnarray}
that are traceless, symmetric and divergence-free in the indices
$\alpha=\alpha_1\cdots\alpha_n$ and the $u$ and $v$ spinor tensors satisfy
\begin{eqnarray}
    \gamma_{\alpha_i} u^{\alpha_1\cdots\alpha_i\cdots\alpha_n} (k,\pm s)
&=& \gamma_{\alpha_i} v^{\alpha_1\cdots\alpha_i\cdots\alpha_n} (k,\pm s)
\, =\,\, 0\,,
\label{eq:gamma_spinor_equation}\\
    \not\!{k}\,u^{\alpha_1\cdots\alpha_i\cdots\alpha_n}(k,\pm s)
&=&\not\!{k}\,v^{\alpha_1\cdots\alpha_i\cdots\alpha_n}(k,\pm s)
\,\,\,\, =\, \,\, 0\,,
\label{eq:massless_on-shell_equation}
\end{eqnarray}
with $\not\!{k}=k_\mu \gamma^\mu$. We adopt the chiral
representation for the Dirac gamma matrices $\gamma^\mu$ ($\mu=0,1,2,3$),
whose expressions are listed in
Appendix~\ref{appendix:scalar_vector_tensor_currents_jacob_wick}.\s

Interchanging two identical massless fermions, i.e taking the opposite
fermion flow line~\cite{Denner:1992vza,Denner:1992me}, we can rewrite
the helicity amplitude of the decay $X\to ff$ with a massless
fermion $M=f$ as
\begin{eqnarray}
    \tilde{\cal M}^{X\to ff}_{\sigma; \lambda_1,\lambda_2}
&=& \bar{u}_2^{\beta}(k_2,\lambda_2)\,\,
    \Gamma^{\mu}_{\beta,\alpha}(p,-k)
     \,\, v_1^{\alpha}(k_1,\lambda_1)\,\,
    \epsilon_{\mu}(p,\sigma)
    \nonumber\\
&=& v_1^{T\alpha }(k_1,\lambda_1)\,\,
     \Gamma^{\mu T}_{\beta,\alpha}(p,-k)
     \,\, \bar{u}_2^{T\beta}(k_2,\lambda_2)\,\,
    \epsilon_{\mu}(p,\sigma)
\label{eq:xmm_vertex_interchanged}
\end{eqnarray}
with the superscript $T$ denoting the transpose of the matrix.
Introducing the charge-conjugation operator $C$ satisfying
$C^\dagger=C^{-1}$ and $C^T=-C$ relating the $v$ spinor to
the $u$ spinor as
\begin{eqnarray}
v^\alpha(k,\lambda) =  C\bar{u}^{T\alpha}(k, \lambda)\,,
\end{eqnarray}
with $\bar{u}=u^\dagger \gamma^0$, we can rewrite the amplitude
as
\begin{eqnarray}
    \tilde{\cal M}^{X\to ff}_{\sigma; \lambda_1,\lambda_2}
&=& - \bar{u}_1^{\beta}(k_1,\lambda_1)\,\,
     C\, \Gamma^{\mu T}_{\beta,\alpha}(p,-k)\, C^{-1}
     \,\, v_2^{\alpha}(k_2,\lambda_2)\,\,
    \epsilon_{\mu}(p,\sigma)\,.
\label{eq:xmm_vertex_interchanged_rewritten}
\end{eqnarray}
Since Fermi statistics requires $\tilde{\cal M}=- {\cal M}$, the triple vertex
tensor must satisfy the relation
\begin{eqnarray}
  C\, \Gamma^{\mu T}_{\beta,\alpha}(p,-k)\, C^{-1}
\, =\,  \Gamma^{\mu}_{\alpha,\beta}(p,k)\,,
\label{eq:fermi_symmetry_relation}
\end{eqnarray}
which enables us to classify all the allowed terms
systematically~\cite{Kayser:1982br,Kayser:1984ge,Boudjema:1990st}.\s

The basic relation for the charge-conjugation invariance of the
Dirac equation is $C\,\gamma^T_\mu\, C^{-1} =- \gamma_\mu$ with a
unitary matrix $C$. Repeatedly using the basic relation, we can derive
\begin{eqnarray}
  \Gamma^c\, \equiv\, C\, \Gamma^T\, C^{-1}
 \,=\, \epsilon_C\,\Gamma \quad
 \mbox{with}\quad
 \epsilon_C
 = \left\{\begin{array}{lll}
        +1  & \mbox{for} & \Gamma = 1, \gamma_5, \gamma_\mu\gamma_5 \\[2mm]
        -1  & \mbox{for} & \Gamma = \gamma_\mu
          \end{array}\right.\,,
\label{eq:charge_conjugation_on_gamma_structure}
\end{eqnarray}
There are no further independent terms as any other operator can be
replaced by a linear combination of $1,\, \gamma_5,\, \gamma_\mu$,
and $\gamma_\mu\gamma_5$ by use of the so-called Gordon identities,
when coupled to the $u$ and $v$ spinors.\s

Because of the conditions (\ref{eq:gamma_spinor_equation}) and
(\ref{eq:massless_on-shell_equation}), the vector structure $\gamma^\mu$
can be contracted only with the wave vector $\epsilon^\mu(p,\sigma)$ of
the decaying particle $X$ with the helicity $\sigma=\pm 1, 0=\pm, 0$.
Effectively we can make the following replacements
\begin{eqnarray}
  S^-_{\alpha\beta}\,\,[1, \gamma_5]
\ \ & \longleftrightarrow & \ \
  -S^+_{\alpha\beta}\,\,[\gamma_5, 1]\,, \\
  \gamma_{\mu_1}\, [1, \gamma_5]\,\, T^-_{\alpha\beta,\mu_2\mu_3}
\ \ & \longleftrightarrow & \ \
  \gamma_{\mu_1}\,[\gamma_5, 1]\,\, T^+_{\alpha\beta,\mu_2\mu_3}\,,
\label{eq:scalar_tensor_replacements}
\end{eqnarray}
because two corresponding terms in each equation give rise to the
same helicity amplitudes as shown explicitly in
Appendix~\ref{appendix:scalar_vector_tensor_currents_jacob_wick}.
Then, for a massless
fermion $M=f$, the general form of the $Xff$ triple vertex tensor can
be written as
\begin{eqnarray}
&&   \Gamma^{\tiny \mbox{$X\to ff$}}_{\mu;\, \alpha,\beta}(p,k)
 = \eta_+\, (x^+_f + x^-_f\, \gamma_5) \,\, \, S^+_{\alpha_1\beta_1}\, \cdots
              S^+_{\alpha_n\beta_n}\,\,
              k_{\mu_1}\,\cdots\, k_{\mu_{_J}}/m^J
    \nonumber\\
&& \mbox{ }\hskip 0.5cm  + \eta_+\,\theta(J-2s)\,\,
     y^+_f \gamma_{\mu_1}\, T^+_{\alpha_1\beta_1,\mu_2\mu_3}
     \,\cdots T^+_{\alpha_n\beta_n,\mu_{2n}\mu_{2n+1}}\,\,
             k_{\mu_{2n+2}}\,\cdots\, k_{\mu_{_J}}/m^{J-2s-1}
    \nonumber\\
&& \mbox{ }\hskip 0.5cm + \eta_-\,\theta(J-2s)\,\,
      y^-_f \gamma_{\mu_1}\gamma_5\,
      T^+_{\alpha_1\beta_1,\mu_2\mu_3}\, \cdots\,
      T^+_{\alpha_n\beta_n,\mu_{2n}\mu_{2n+1}}\,\,
      k_{\mu_{2n+2}}\cdots\, k_{\mu_{_J}}/m^{J-2s-1} \,,
\label{eq:general_x_ff_vertex_tensor}
\end{eqnarray}
in terms of four independent parameters, $x^\pm_f$ and $y^\pm_f$,
with $n=s-1/2$. The scalar part proportional to
$g_{_{\bot}\alpha_i\beta_i} k_{\mu_{2i}}k_{\mu_{2i+1}}$
of  $T^+_{\alpha_i\beta_i,\mu_{2i}\mu_{2i+1}}$ in
Eq.$\,$(\ref{eq:scalar_tensor_operators_3})
can be discarded because its contribution
vanishes in the opposite helicity case with $\lambda_1=-\lambda_2$
enforced by the helicity-preserving (axial-)vector currents.
It is straightforward to check that the expression
(\ref{eq:general_x_ff_vertex_tensor})
satisfies the Fermi symmetry
condition (\ref{eq:fermi_symmetry_relation}) for two identical
fermions. One non-trivial observation is that a spin-1
particle can decay only into two identical spin-1/2
particles~\cite{Boudjema:1990st} through an axial-vector $\gamma_\mu \gamma_5$
current as can be checked with the last expression
in Eq.$\,$(\ref{eq:general_x_ff_vertex_tensor}).\s

\section{Explicit form of reduced helicity amplitudes}
\label{sec:explicit_form_reduced_helicity_amplitudes}

Employing all the analytic results for the scalar, vector and tensor
currents listed
in Appendix~\ref{appendix:scalar_vector_tensor_currents_jacob_wick}
enables us to explicitly calculate all the reduced helicity amplitudes
${\cal C}_{\lambda_1,\lambda_2}$ defined
in Eq.$\,$(\ref{eq:x_mm_helicity_amplitude}) both in the
integer-spin and half-integer-spin cases for the massless
particle.\s

{\it In the case with a massless integer-spin $s$ boson $M=b$},
the reduced helicity amplitudes for the process $X\to bb$
with the same helicities $(\pm,\pm)$  read
\begin{eqnarray}
 {\cal C}_{\pm,\pm}
= \frac{2^{J/2}J!}{\sqrt{(2J)!}}\,\,
  (x^+_b \pm x^-_b)\,,
\end{eqnarray}
surviving only for even-integer $J$. If parity is preserved, the term, $x^+_b$
or $x^-_b$, can survive for the even or odd $X$ intrinsic parity
$\eta=\pm$, respectively.
On the other hand, the reduced helicity amplitudes with the opposite helicities
$(\pm,\mp)$, which survive only when $J\geq 2s$, read
\begin{eqnarray}
 {\cal C}_{\pm,\mp}
= 2^{J/2} \sqrt{\frac{(J+2s)!\, (J-2s)!}{(2J)!}}\,\, y^+_b\,,
\end{eqnarray}
for even-integer spin $J\geq 2s$ and even $X$ intrinsic parity
and
\begin{eqnarray}
{\cal C}_{\pm,\mp}
= \mp 2^{J/2} \sqrt{\frac{(J+2s)!\, (J-2s)!}{(2J)!}}\,\, y^-_b\,,
\end{eqnarray}
for odd-integer spin $J\geq 2s$ and odd $X$ intrinsic parity.\s

{\it In the case with a massless fermion $M=f$} with a half-integer
spin $s$, the reduced helicity amplitudes for the process $X\to ff$ with
the same helicities $(\pm,\pm)$ read
\begin{eqnarray}
  {\cal C}_{\pm,\pm}
= \frac{2^{J/2}J!}{\sqrt{(2J)!}}\,\, (x^+_f \mp x^-_f)\,,
\end{eqnarray}
surviving only for even-integer $J$. If parity is preserved, the $x^+_f$
and $x^-_f$ terms can survive for the odd and even $X$
intrinsic parity $\eta=\mp$, respectively.
On the other hand, the reduced helicity amplitudes, which survive only
when $J\geq 2s$, read
\begin{eqnarray}
{\cal C}_{\pm,\mp}
= (-1)^{2s}\, 2^{J/2} \sqrt{\frac{(J+2s)!\, (J-2s)!}{(2J)!}}\,\,
  y^+_f\,,
\end{eqnarray}
for even-integer spin $J\geq 2s$ and odd $X$ intrinsic parity $\eta=-$,
and
\begin{eqnarray}
{\cal C}_{\pm,\mp}
= \mp (-1)^{2s}\, 2^{J/2} \sqrt{\frac{(J+2s)!\, (J-2s)!}{(2J)!}}\,\,
  y^-_f\,,
\end{eqnarray}
for odd-integer spin $J\geq 2s$ and even $X$ intrinsic parity $\eta=+$.\s

\section{Conclusions}
\label{sec:conclusions}

We have generalized the well-known Landau-Yang (LY) theorem
on the decay of a neutral particle into two photons for analyzing the two-body
decay of a neutral or charged particle into two identical massless particles
of any integer or half-integer spin. After having worked out the selection 
rules classified by discrete parity invariance and Bose/Fermi symmetry in
the helicity formulation, we have derived the general form of
the Lorentz covariant triple vertices and calculated the helicity amplitudes
explicitly.  After checking the consistency of all the analytic
results obtained from two complementary approaches, we have
drawn out the key aspects of the generalized LY theorem including the
re-interpretation of the WW theorem.\s

Introducing a compact notation $n\, [J, s]^{\zeta}$ consisting of the
number $n$ of independent terms, the $X$ and $M$ spins, $J$ and $s$, and
a parity $\zeta=\eta (-1)^{J-2s}$ with the $X$ intrinsic parity $\eta$,
we can summarize the selection rules on the decay of a massive particle
into an identical pair of massless particles of arbitrary spin:
\begin{eqnarray}
      n\, [J, s]^{\zeta}
&=& 1\, [0, 0]^+\,,\ \ 1\, [0,s>0 ]^\pm\,, \ \
    1\, [2k-1, 0<s\leq k-1/2]^-\,,
       \nonumber\\
&&  1\, [2k, 0]^+\,, \ \
       2\, [2k, 0<s \leq k]^+\,, \ \
       1\, [2k, 0<s \leq k]^-\,, \ \
       1\, [2k, s > k ]^\pm\,,
\end{eqnarray}
with $k=1,2,\cdots$. In the special case of $s=1$, the selection rules reduce
to those called the LY theorem described in
Eq.$\,$(\ref{eq:landau_yang_theorem}). The massless particle
for the decay of a spin-1 particle into its identical pair must
have a half-integer spin $s=1/2$. If parity is preserved, no odd-spin
and odd-parity (even-parity) particle can decay into an identical pair of
massless bosons (fermions). \s

As natural extension of the present work, the decays of an arbitrary
integer-spin particle into two identical {\it massive}
particles~\cite{Boudjema:1990st} or two distinguishable charge self-conjugate
particles of any spin are presently under study and its results will be
reported separately. Before closing, we emphasize that the selection
rules obtained in this letter can be applied without any modifications,
for example, to the decay of a doubly-charged massive particle into two
identical singly-charged massless particles, where the charge can be of
any type as well as of a typical electric charge type.  \s

\section*{Acknowledgments}
\label{sec:acknowledgments}

The work was in part by the Basic Science Research Program of Ministry of
Education through National Research Foundation of Korea
(Grant No. NRF-2016R1D1A3B01010529) and in part by the CERN-Korea theory
collaboration. A correspondence with Kentarou Mawatari
concerning the covariant formulation of the Landau-Yang theorem
is acknowledged.\s

\appendix

\section{Wave vectors and spinors in the Jacob-Wick convention}
\label{appendix:wave_vectors_spinors_jacob_wick}

In Appendix~\ref{appendix:wave_vectors_spinors_jacob_wick},
we show the explicit expressions for the wave vectors and spinors
of a massive particle $X$ and two massless particles in the $X$RF
with the kinematic configuration as shown
in Figure~\ref{fig:kinematic_configuration_xmm_xrf}. The Jacob-Wick 
convention
of Ref.$\,$\cite{Jacob:1959at} is chosen for the vectors and spinors.
In terms of the polar and azimuthal angles, $\theta$ and $\phi$,
the three momenta, $p=k_1+k_2$ and $k_{1,2}$, and the combination
$k=k_1-k_2$ read
\begin{eqnarray}
p \,\, &=& m\, (1, 0,0,0)\,, \\
k_1 &=& \frac{m}{2}\,
        (1, \phantom{+}\sin\theta\cos\phi, \phantom{+}\sin\theta\sin\phi,
         \phantom{+}\cos\theta)\,,\\
k_2 &=& \frac{m}{2}\,
        (1, -\sin\theta\cos\phi, -\sin\theta\sin\phi, -\cos\theta)\,, \\
k\,\, &=& m\, (0, \phantom{+}\sin\theta\cos\phi, \phantom{+}\sin\theta\sin\phi,
        \phantom{+}\cos\theta)\,,
\end{eqnarray}
For the sake of notation, we introduce three unit vectors
\begin{eqnarray}
   \hat{k}
&=& (\sin\theta\cos\phi,\, \sin\theta\sin\phi,\, \phantom{+}\cos\theta)
\, =\, \vec{k}/m\,, \\
   \hat{\theta}
&=& (\cos\theta\cos\phi,\, \cos\theta\sin\phi,\, -\sin\theta) \,, \\
   \hat{\phi}
&=& (-\sin\phi,\, \cos\phi,\, 0)\,,
\end{eqnarray}
which are mutually orthonormal, i.e. $\hat{k}\cdot\hat{\theta}=\hat{\theta}\cdot
\hat{\phi}=\hat{\phi}\cdot\hat{k}=0$ and $\hat{k}\cdot\hat{k}=\hat{\theta}\cdot
\hat{\theta}=\hat{\phi}\cdot\hat{\phi}=1$.\s

The wave vectors for the particle with momentum $p$ and two massless
particles with momenta $k_{1,2}$ are given by
\begin{eqnarray}
\epsilon(p,\pm) &=& \frac{1}{\sqrt{2}}\, (0,\, \mp 1,\,  -\imath,\, 0)\,,
     \label{eq:x_pm_wave_vector} \\
\epsilon(p,\, 0)  &=& (0,\, 0,\, 0,\, 1)\,,
     \label{eq:x_0_wave_vector} \\
\epsilon_1(k_1,\pm) &=& \frac{1}{\sqrt{2}}\, e^{\pm \imath \phi}\,
        (0,\, \mp \cos\theta\cos\phi+\imath \sin\phi,\,
              \mp \cos\theta\sin\phi-\imath \cos\phi,\,
              \pm \sin\theta)\nonumber\\
        &=& \frac{1}{\sqrt{2}}\, e^{\pm \imath \phi}\,
        (0, \mp \hat{\theta}-\imath \hat{\phi})\,,
     \label{eq:m1_pm_wave_vector} \\
\epsilon_2(k_2,\pm) &=& \epsilon_1(k_1, \mp)
                  \, = \, -\epsilon^*_1(k_1,\pm)
                  \, = \, -\epsilon^*_2(k_2,\mp)\,,
     \label{eq:m2_pm_wave_vector}
\end{eqnarray}
where the last relation between two wave vectors is satisfied in
the Jacob-Wick convention. \s

The spin-1/2 $u$ and $v$ 4-component spinors of the massless particles
with momenta $k_{1,2}$ are given in the Jacob-Wick convention by
\begin{eqnarray}
&& u_1(k_1,+)\, = -v_1(k_1, -)= \sqrt{m}\, \left(\begin{array}{c}
                           0 \\[1mm]
                           \chi_+(\hat{k})
                   \end{array}\right)\,, \\
&& u_1(k_1,-)\, = -v_1(k_1,+)= \sqrt{m}\, \left(\begin{array}{c}
                           \chi_-(\hat{k})\\[1mm]
                           0
                   \end{array}\right)\,, \\
&& u_2(k_2,+)\, = \phantom{+} v_2(k_2,-)
            = \sqrt{m}\, \left(\begin{array}{c}
                           0 \\[1mm]
                           \chi_-(\hat{k})
                   \end{array}\right)\,, \\
&& u_2(k_2,-)\, = \phantom{+} v_2(k_2,+)= \sqrt{m}\, \left(\begin{array}{c}
                           \chi_+(\hat{k})\\[1mm]
                           0
                   \end{array}\right)\,,
\end{eqnarray}
satisfying the relations $u_2(k_2,\pm)=\gamma^0 u_1(k_1,\mp)$ and
$v_2(k_2,\pm)=-\gamma^0 v_1(k_1,\mp)$ with the expression of $\gamma^0$ given
in Appendix~\ref{appendix:scalar_vector_tensor_currents_jacob_wick}, where
the 2-component spinors $\chi_\pm(\hat{k})$ are written in terms of
the polar and azimuthal angles, $\theta$ and $\phi$, as
\begin{eqnarray}
\chi_+(\hat{k}) &=& e^{i\phi/2}\,\left(\begin{array}{l}
                       \cos\frac{\theta}{2}\, e^{-i\phi/2}\\[1mm]
                       \sin\frac{\theta}{2}\, e^{i\phi/2}
                       \end{array}\right)\,, \\
\chi_-(\hat{k}) &=& e^{-i\phi/2}\left(\begin{array}{c}
                       -\sin\frac{\theta}{2}\, e^{-i\phi/2} \\[1mm]
                       \!\!\!\!\!\!\cos\frac{\theta}{2} \, e^{i\phi/2}
                       \end{array}\right)\,,
\end{eqnarray}
being mutually orthonormal, i.e. $\chi^\dagger_a(\hat{k})
\chi_b(\hat{k})=\delta_{a,b}$, with $a,b=\pm$.\s

\section{Scalar, vector and tensor currents in the Jacob-Wick convention}
\label{appendix:scalar_vector_tensor_currents_jacob_wick}

Appendix~\ref{appendix:scalar_vector_tensor_currents_jacob_wick} contains the
list of the expressions for several scalar, vector and tensor currents in
the kinematic configuration depicted in
Figure~\ref{fig:kinematic_configuration_xmm_xrf} to be used
in the main text. For explicit current calculations, the
chiral representation is adopted for four anti-commutating Dirac gamma
matrices $\gamma^\mu$ with $\mu=0,1,2,3$ and $\gamma_5=\imath
\gamma^0\gamma^1\gamma^2\gamma^3$, whose expressions are given by the
following $4\times 4$ matrices
\begin{eqnarray}
\gamma^\mu = \left(\begin{array}{cc}
                   0 & \sigma^\mu_+ \\[2mm]
                   \sigma^\mu_- & 0
                   \end{array}\right)
           \quad\mbox{and}\quad
\gamma_5 = \left(\begin{array}{cc}
                   -1 & 0 \\[2mm]
                    0 & 1
                   \end{array}\right)\,,
\label{eq:gamma_matrices_chiral_representation}
\end{eqnarray}
with $\sigma^\mu_\pm=(1,\,\, \pm \vec{\sigma})$ in terms of three Pauli
matrices $\vec{\sigma}=(\sigma_1,\sigma_2,\sigma_3)$ defined by
\begin{eqnarray}
    \sigma_1
\, =\,  \left(\begin{array}{cc}
           0  & 1 \\[2mm]
           1  & 0
           \end{array}\right)\,,\quad
               \sigma_2
\, =\,  \left(\begin{array}{cc}
           0  & -\imath  \\[2mm]
           \imath   & 0
           \end{array}\right)\,,\quad
               \sigma_3
\, =\, \left(\begin{array}{cc}
           1  & 0 \\[2mm]
           0  & -1
           \end{array}\right)\,.
\end{eqnarray}
The metric tensor is defined to be $g^{\mu\nu}=g_{\mu\nu}
={\rm diag}(1,-1,-1,-1)$.\s

Firstly, the helicity-dependent scalar and pseudo-scalar currents,
surviving only when two helicities are same, i.e. $\lambda_1=\lambda_2$, 
read
\begin{eqnarray}
  u^+_{\lambda_1\lambda_2}
&\equiv& \bar{u}_1(k_1,\lambda_1)v_2(k_2,\lambda_2)/m
\ \ \, \, =\,  \phantom{+} \delta_{\lambda_1,\lambda_2}\,,\\
 u^+_{\lambda_1\lambda_2}
&\equiv&
 \bar{u}_1(k_1,\lambda_1) \gamma_5 v_2 (k_2,\lambda_2)/m
\, =\, - \lambda_1\delta_{\lambda_1,\lambda_2}\,,
\label{eq:scalar_pseudoscalar_currents}
\end{eqnarray}
where $\lambda_{1,2}=\pm 1 =\pm$ which is two times the helicity,
$\pm 1/2$.
Secondly, the helicity-dependent vector and axial-vector currents,
surviving only when two helicities are opposite, i.e. $\lambda_1=-\lambda_2$,
read
\begin{eqnarray}
 \bar{u}_1(k_1,\lambda_1)\gamma^\mu v_2 (k_2,\lambda_2)
&=& -\sqrt{2}\, m\,\,\,\, \delta_{\lambda_1,-\lambda_2}\,\,\,\,\,
     \epsilon^{*\mu}_1(k_1,\lambda_1)\,,\\
 \bar{u}_1(k_1,\lambda_1)\gamma^\mu \gamma_5 v_2(k_2,\lambda_2)
&=& -\sqrt{2}\, m\, \lambda_1 \delta_{\lambda_1,-\lambda_2}\,\,\,
      \epsilon^{*\mu}_1(k_1,\lambda_1)\,,
\label{eq:vector_axialvector_currents}
\end{eqnarray}
with $\lambda_{1,2}=\pm 1=\pm$ where the expression of the wave
vector $\epsilon^\mu_1(k_1,\lambda_1)$ is given in
Eq.$\,$(\ref{eq:m1_pm_wave_vector}).\s

The contraction of the $X$ wave vector $\epsilon(p,\sigma)$ with
$\sigma=\pm, 0$ and $k$ gives
\begin{eqnarray}
    k\cdot\epsilon(p,\, 0)\,
&=& -m\cos\theta\,, \\
    k\cdot\epsilon(p,\pm)
&=& \pm m \frac{1}{\sqrt{2}} \sin\theta\, e^{\pm i \phi}\,,
\end{eqnarray}
in terms of the polar and azimuthal angles, $\theta$ and $\phi$, in the $X$RF.\s

For the sake of efficiently calculating and deriving the (reduced)
helicity amplitudes of the decay $X\to MM$, we introduce four
helicity-dependent quantities, $s^\pm_{\lambda_1\lambda_2}$
and $t^\pm_{\lambda_1\lambda_2}$, defined by contracting the scalar
and tensor operators given in Eqs.$\,$(\ref{eq:scalar_tensor_operators_1}),
(\ref{eq:scalar_tensor_operators_2}), (\ref{eq:scalar_tensor_operators_3})
and (\ref{eq:scalar_tensor_operators_4})
with the $X$ and $M$ wave vectors appropriately as
\begin{eqnarray}
  s^\pm_{\lambda_1\lambda_2}
&=& \epsilon_1^{*\alpha}(k_1,\lambda_1)\,
    \epsilon_2^{*\beta}(k_2,\lambda_2)\,
    S^\pm_{\alpha\beta}\,\,,
\label{eq:sclar_components}\\
  t^\pm_{\lambda_1\lambda_2}
&=& \epsilon_1^{*\alpha}(k_1,\lambda_1)\,
    \epsilon_2^{*\beta}(k_2,\lambda_2)\,
    T^\pm_{\alpha\beta,\mu\nu}\,
    \epsilon^\mu(p,+)\,
    \epsilon^\nu(p,+)\,,
\label{eq:tensor_components}
\end{eqnarray}
and two additional helicity-dependent quantities obtained by contracting
the vector and axial-vector currents with the $X$ wave vector
$\epsilon(p,+)$ as
\begin{eqnarray}
  v^+_{\lambda_1\lambda_2}
&=& \bar{u}_1(k_1,\lambda_1)\,\,\gamma_\mu\,\, v_2(k_2,\lambda_2)\,\,
    \epsilon^\mu(p,+)/m\,,
\label{eq:fermion_+_component} \\
  v^-_{\lambda_1\lambda_2}
&=& \bar{u}_1(k_1,\lambda_1)\gamma_\mu\gamma_5 v_2(k_2,\lambda_2)\,
    \epsilon^\mu(p,+)/m\,,
\label{eq:fermion_+_component}
\end{eqnarray}
with $\lambda_{1,2}=\pm 1=\pm$. Explicitly the
same-helicity quantities and the opposite-helicity quantities read
\begin{eqnarray}
  s^+_{\pm\pm}
&=& \pm\, s^-_{\pm\pm}\, =\, 1 \,,
\label{eq:explicit_boson_case_1}\\
  t^+_{+-}
&=&  \phantom{+} (1+\cos\theta)^2/2 \hskip 1.4cm
  \, =\,  2\, d^2_{2,2}(\theta)\,,
\label{eq:explicit_boson_case_2}\\
  t^+_{-+}
&=& \phantom{+} (1-\cos\theta)^2\, e^{4\imath \phi}/2\hskip 0.7cm
  \, =\, 2\, d^2_{2,-2}(\theta)\, e^{4\imath \phi}\,,
\label{eq:explicit_boson_case_3}\\
  t^-_{\pm\mp}
&=&  \pm t^+_{\pm\mp}\,,
\label{eq:explicit_boson_case_4}
\end{eqnarray}
and the quantities involving the spinors in the fermionic case
\begin{eqnarray}
  u^+_{\pm\pm}
&=& \mp u^-_{\pm\pm} = 1\,,
  \label{eq:explicit_fermion_case_1}\\
  v^+_{+-}
&=& \phantom{+} (1+\cos\theta)/\sqrt{2}
\hskip 1.2cm
 \, =\, \sqrt{2}\, d^1_{1,1}(\theta)\,,
  \label{eq:explicit_fermion_case_2}\\
  v^+_{-+}
&=& \phantom{+} (1-\cos\theta)\, e^{2\imath \phi}/\sqrt{2}
 \hskip 0.5cm
 \, =\, \sqrt{2}\, d^1_{1,-1}(\theta)\, e^{2\imath \phi}\,,
  \label{eq:explicit_fermion_case_3}\\
 v^-_{\pm\mp}
&=& \pm v^+_{\pm\mp}\,,
  \label{eq:explicit_fermion_case_4}
\end{eqnarray}
in terms of the polar and azimuthal angles $\theta$ and $\phi$ in the
$X$RF. All the quantities with other helicity combinations are zero.\s

The validity of the replacements in
Eq.$\,$(\ref{eq:scalar_tensor_replacements}) can be confirmed with the
four relations
\begin{eqnarray}
 s^+_{\pm\pm} u^\pm_{\pm\pm}
&=& -\, s^-_{\pm\pm} u^\mp_{\pm\pm}\,, \\
  t^+_{\pm\mp} v^\pm_{\pm\mp}
&=& +\, t^-_{\pm\mp} v^\mp_{\pm\mp}\,,
\end{eqnarray}
which can be checked by use of all the expressions from
Eq.$\,$(\ref{eq:explicit_boson_case_1})
to Eq.$\,$(\ref{eq:explicit_fermion_case_4}). Moreover, in the general and
covariant form, we have the corresponding four identities for non-zero
decay helicity amplitudes
\begin{eqnarray}
&&    S^-_{\alpha\beta}\,\,
      \epsilon_1^{*\alpha}(k_1,\pm)\,\epsilon_2^{*\beta}(k_2,\pm)\,\,
      \bar{u}_1(k_1,\pm)\,[1, \gamma_5]\, v_2(k_2,\pm) \nonumber\\
&& \mbox{ }\hskip 2.cm
\, =\,  - S^+_{\alpha\beta}\,\,
     \epsilon_1^{*\alpha}(k_1,\pm)\,\epsilon_2^{*\beta}(k_2,\pm)\,\,
     \bar{u}_1(k_1,\pm)\, [\gamma_5, 1]\, v_2(k_2,\pm)\,,
\label{eq:effective_replacements_1}\\
&&   T^-_{\alpha\beta,\mu_2\mu_3}\,
     \epsilon_1^{*\alpha}(k_1,\pm)\,\epsilon_2^{*\beta}(k_2,\mp)\,\,
      \bar{u}_1(k_1,\pm)\,\gamma_{\mu_1}\, [1, \gamma_5]\, v_2(k_2,\mp)\,\,
      \nonumber \\
&& \mbox{ }\hskip 2.cm
   \,=\,  \phantom{+} T^+_{\alpha\beta,\mu_2\mu_3}\,
     \epsilon_1^{*\alpha}(k_1,\pm)\,\epsilon_2^{*\beta}(k_2,\mp)\,\,
     \bar{u}_1(k_1,\pm)\, \gamma_{\mu_1}\, [\gamma_5, 1]\, v_2(k_2,\mp)\,,
\label{eq:effective_replacements_2}
\end{eqnarray}
where the last term $-g_{_{\bot}\alpha\beta}\, k_{\mu_2} k_{\mu_3}/m^2$ of
the tensor $T^+_{\alpha\beta,\mu_2\mu_3}$ defined
in Eq.$\,$(\ref{eq:scalar_tensor_operators_3}) does not contribute to
the expression on the right-hand side of
Eq.$\,$(\ref{eq:effective_replacements_2}) surviving only in the
opposite-helicity case with $\lambda_1=-\lambda_2=\pm$.\s

The angle-dependent parts are encoded solely in the Wigner-$d$
functions and the reduced helicity amplitudes are independent of
the helicity of the decaying particle. Therefore, it is sufficient to
consider the cases with the maximal $X$ helicity $\sigma=J$
and the zero and two maximal helicity differences,
$\lambda_1-\lambda_2=0,\pm 2s$ for deriving the reduced helicity 
amplitudes. For them, three relevant Wigner-$d$ functions read
\begin{eqnarray}
    d^{J}_{J,\, 0}(\theta)
&=& \frac{(-1)^J}{2^J}\, \frac{\sqrt{(2J)!}}{J!}\, \sin^J\theta\,, \\
    d^{J}_{J,\, 2s} (\theta)
&=& \frac{(-1)^{J-2s}}{2^J}
    \sqrt{\frac{(2J)!}{(J+2s)!\,(J-2s)!}}
    \, (1+\cos\theta)^{2s}\, \sin^{J-2s}\theta\,,\\
  d^{J}_{J,-2s}(\theta)
&=& \frac{(-1)^{J+2s}}{2^J}
    \sqrt{\frac{(2J)!}{(J+2s)!\,(J-2s)!}}
    \, (1-\cos\theta)^{2s}\, \sin^{J-2s}\theta\,,
\end{eqnarray}
which are to be factored out for deriving the reduced helicity amplitudes
in Section~\ref{sec:explicit_form_reduced_helicity_amplitudes}.\s

\end{document}